\documentclass[twocolumn,aps,prl]{revtex4}
\usepackage{graphics}
\usepackage{graphicx}
\usepackage{dcolumn}
\usepackage{amsmath}
\usepackage{latexsym}
\usepackage[utf8]{inputenc}
\begin {document}

\title {Brittle to quasibrittle transition in a compound fiber bundle}
\author
{
Chandreyee Roy and S. S. Manna
}
\affiliation
{
\begin {tabular}{c}
Satyendra Nath Bose National Centre for Basic Sciences,
Block-JD, Sector-III, Salt Lake, Kolkata-700106, India \\
\end{tabular}
}
\begin{abstract}
      The brittle to quasibrittle transition has been studied for a compound of two different kinds of fibrous 
   materials, having distinct difference in their breaking strengths under the framework of the fiber bundle model.
   A random fiber bundle model has been devised with a bimodal distribution of the 
   breaking strengths of the individual fibers. The bimodal distribution is assumed to be consisting of two 
   symmetrically placed rectangular probability distributions of strengths $p$ and $1 - p$, 
   each of width $d$, and separated by a gap $2s$. Different properties of the transition have been studied 
   varying these three parameters and using the well known equal load sharing dynamics. Our study exhibits a
   brittle to quasibrittle transition at the critical width $d_c(s,p) = p(1/2 - s)/(1 + p)$ confirmed by our
   numerical results.
\end{abstract}
\maketitle

\section{I. Introduction}

      Understanding the behavior and properties of different materials subjected to applied external stress 
   is important for their useful applications as well as for the prevention of their mechanical failures. 
   The fiber bundle model (FBM) is a simple framework to study such breakdown processes where the material 
   is in the form of a bundle of thin fibers. This model was introduced by Pierce \cite{Pierce} to study 
   the strength of cotton yarns which was then further extended from the point of view of statistical 
   physics by Daniels \citep{Daniels}. Moreover, the catastrophic global failures in FBMs have also been 
   studied in the context of earthquakes, traffic systems etc. 
   \cite{RoyHatano,Pradhan0,KunZapperi,Pradhan1,HansenBook,Pradhan2,Herrmann,Chakrabarti,
   Sornette,Sahimi,Bhattacharya}. In this paper we have studied the statistical properties of FBM when 
   the bundle consists of a mixture of two different types of fibers, e.g., cotton and nylon.
\begin{figure}[t]
\includegraphics[width=6.0cm]{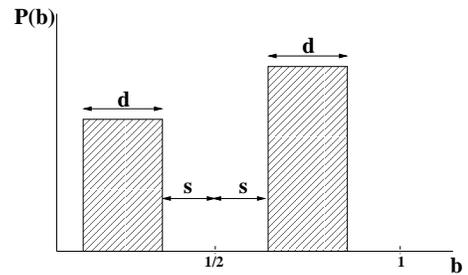}
\caption{
The bimodal probability density function $P(b)$ of the breaking thresholds $b$ of the fibers plotted against $b$. 
The distribution consists of two rectangular blocks of width $d$, symmetrically placed around $b=1/2$ keeping a gap 
of $2s$ between them. The total probabilities in the first and the second blocks are denoted by $p$ and $1 - p$ respectively.
}
\label{FIG01}
\end{figure}

      In FBM, a set of $N$ fibers is placed in parallel to one another. These fibers are imagined to be 
   thin elastic massless strings suspended vertically, clamped at the lower end and supported rigidly 
   at the top. An external load $F$ is applied to the entire bundle at the bottom to stretch 
   the fibers. Every fiber $i$ has a distinct breaking threshold $b_i$ of its own whose value is
   randomly drawn from a probability density function $P(b)$. 

      In this model, the failure process follows a stress conservative dynamics. Under the applied stress,
   each fiber elongates linearly obeying the Hooke's law. 
   For simplicity, the Young's modulus for every fiber is assumed to be equal to unity. On increasing the 
   load the stress acting through the $i$-th fiber reaches its breaking threshold $b_i$, beyond which 
   it fails. In general, when a fiber fails, it releases the stress that was acting through it. The 
   released stress then gets distributed equally among all the remaining intact fibers. This procedure 
   is referred as the Equal load sharing (ELS) dynamics. Consequently, the stresses acting through 
   the intact fibers are increased which may result in the failure of additional fibers though the 
   external load $F$ has been maintained to the same value. Thus, quite often the failure of only a
   single fiber triggers a cascade of fiber failures, known as the `avalanche'. The avalanche
   terminates when a stable state is reached where there is no more fiber failure since each intact 
   fiber has its breaking threshold larger than the externally applied load per fiber. When 
   $F$ is enhanced quasi-statically, the complete failure of the bundle takes place in a sequence
   of such successive avalanches \cite {Hidalgo, Kloster, Pradhan4, Hansen}.

      In general, fiber bundles are classified into two different categories depending on how the 
   entire bundle fails. If the failure of the weakest fiber, having the smallest breaking threshold, 
   results in a huge avalanche which leads to the failure of the entire bundle, the bundle is 
   referred as `brittle'. On the other hand, if more than one avalanche are needed to break all 
   the fibers, the fiber bundle is called `quasibrittle'. Whether a fiber bundle would 
   exhibit brittle or quasibrittle type failure, is determined by the characteristics of the 
   probability density function $P(b)$. Recently, transitions from the 
   brittle to quasibrittle states have been studied for FBMs with linear and non-linear elastic 
   fibers \cite {Subhadeep,Roy3,Kovacs,SRoy}. 
      
      We have studied here such transitions for the compound FBMs.
   Compound materials with mixtures of different kinds of fibers are very 
   important in industrial applications. For example, good quality fabrics are produced using 
   mixtures of cotton and nylon fibers. Breaking thresholds of fibers of these materials are 
   widely different. Keeping this in mind, one therefore likes to ask what changes in 
   the properties of a fiber bundle take place when the bundle is made of two different types 
   of fibers with different sets of breaking thresholds. This prompts us to study the properties 
   of a compound fiber bundle model when the failure thresholds of the individual fibers assume 
   a bimodal distribution. A fiber bundle model with discontinuous distribution of breaking thresholds
   in a different form had been studied in the literature to consider the breakdown properties
   \cite {AmitDutta,AmitDutta1}. It has been shown that introduction of a lower cutoff in the 
   breaking threshold distribution of the fibers affects properties of fiber bundle model 
   \cite{Pradhan3,Raischel}.
      
      In section 2, we describe the characteristics of the bimodal distribution 
   and determine the critical threshold of the compound fiber bundle. In section 3, we describe 
   the brittle to quasibrittle transition in this model of the fiber bundle. Finally, we 
   summarize our work and conclude in section 4.

\begin{figure}[t]
\includegraphics[width=6.0cm]{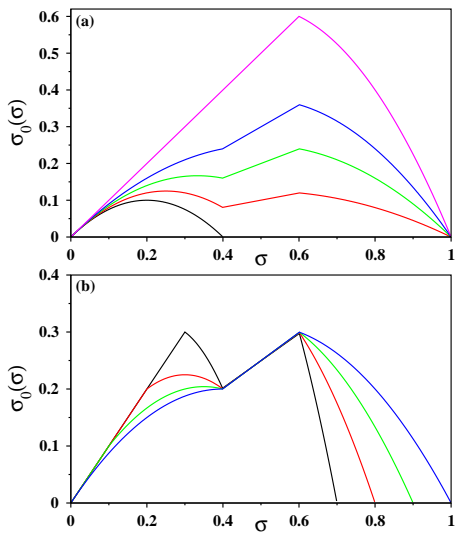}
\caption{Using the Eqn. (\ref{EQN04}), $\sigma_0(\sigma)$ has been plotted against $\sigma$ for 
different parameter values: 
   (a) For $s = 0.1$ and $d = 0.4$ five curves have been plotted for $p$ = 0 (magenta),
   0.4 (blue), 0.6 (green), 0.8 (red) and 1 (black) displayed from top to bottom. 
   (b) For $s$ = 0.1 and $p$ = 0.5 four curves have been plotted for $d$ = 0.1 (black), 0.2 (red), 
   0.3 (green) and 0.4 (blue) displayed from top to bottom in the first block and left to right in 
   the second block.
}
\label{FIG02}
\end{figure}

\section{II. Bimodal distribution and the breaking threshold}

      We consider a fiber bundle having $N$ fibers whose breaking thresholds $\{b_i, i = 1, \ldots, N\}$ are drawn 
   from a bimodal distribution. This distribution is a combination of two uniform distributions 
   of width $d$, symmetrically placed about the midpoint $b = 1/2$ of the $b$ axis, and are separated 
   by an amount of $2s$ as shown in Fig. \ref{FIG01}. The first and the second blocks are 
   extended over the regions $1/2-s-d \leq b \leq 1/2-s$ and $1/2+s \leq b \leq 1/2+s+d$ respectively. 
   The probability that the breaking threshold of a randomly selected fiber is in the first and the 
   second block are denoted by $p$ and $1 - p$ respectively. On the other hand, the probability that
   its breaking threshold lies between $b$ and $b + db$ is denoted by $P(b)db$. Similarly, the cumulative 
   probability $P_c(b)$ that the fiber has strength less than $b$ is given by:
\begin{equation}
P_c(b) = 
\begin{cases} 
p(b -1/2 + s + d)/d,   & \text{for block 1} \\
p + (1-p)(b -1/2-s)/d, & \text{for block 2}.
\end{cases}
\end{equation}

\begin{figure}[t]
\includegraphics[width=6.0cm]{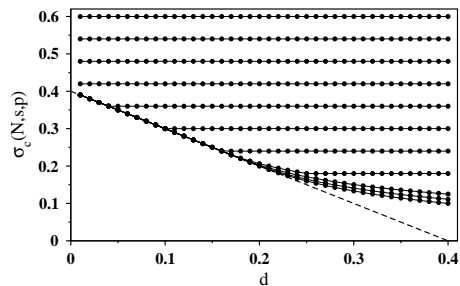}
\caption{
      Plot of the critical load $\sigma_c(N,s,p)$ against the block width $d$ using $N = 2^{18}$ 
   and $s = 0.1$. The value of first block probability $p$ has been tuned between 0 and 1 in steps 
   of 0.1 displayed from top to bottom. The dashed line has the equation 
   $\sigma_c(N,s,p) = 1/2 - s - d$.
}
\label{FIG03}
\end{figure}

      In an arbitrary intermediate stable state the average value of the external load per fiber 
   $\sigma_0(\sigma)$ is given by \cite {Pradhan1,Pradhan2,Roy}
\begin{equation}
\sigma _0(\sigma) = \sigma(1-P_c(\sigma)). 
\end{equation}
   For our compound fiber bundle we have
\begin{equation}
\sigma _0(\sigma) = 
\begin{cases}
[1 - p(\sigma -1/2+s+d)/d]\sigma & \text{for block 1} \\
[(1-p) - (1-p)(\sigma -1/2-s)/d]\sigma &  \text{for block 2}.
\end{cases}
\label{EQN04}
\end{equation}

\begin{figure}[t]
\includegraphics[width=6.0cm]{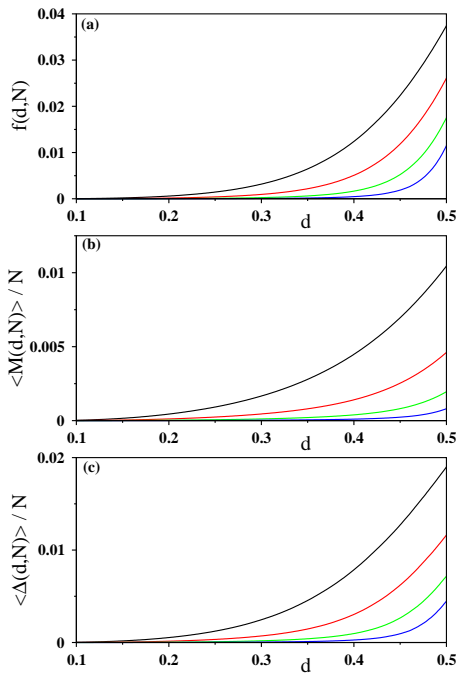}
\caption{
   Plots are shown here for the special case when only the second block exists i.e., $p = 0$.
   The separation parameter $s = 0$ i.e., the left end of the block is fixed at $b = 1/2$ and 
   its right end extends up to $1/2 + d$. Four different bundle sizes are used: $N=2^8$ (black), 
   $2^{10}$ (red), $2^{12}$ (green) and $2^{14}$ (blue) with $N$ increasing from top to bottom. 
   To characterize the brittle to quasibrittle transition, three quantities have been plotted 
   against the block width $d$. They are: (a) the fraction $f(d,N)$ of fibers broken before the 
   last avalanche; (b) the fraction of the average number of avalanches $\langle M(d,N) \rangle / N$ 
   required for complete breakdown and (c) the average avalanche size $\langle \Delta(d,N) \rangle$ 
   scaled by the bundle size.
   }
  \label{FIG04}
\end{figure}

      This variation of $\sigma_0(\sigma)$ against $\sigma$ has been displayed in Fig. 
   \ref{FIG02}(a) for a specific set of values of the parameters $s$ = 0.1, $d$ = 0.4 and for 
   five different values of the first block probability $p$ = 0, 0.4, 0.6, 0.8, and 1. When 
   $p=0$, then all the fibers are in the second block which implies that all the breaking 
   threshold values are confined between $1/2 + s$ and 1. Here, the system is always observed to 
   be brittle and this behavior is evident from the plot in Fig. \ref{FIG02}(a) which shows that 
   $\sigma_0(\sigma)$ is always a decreasing function of $\sigma$. As the value of $p$ 
   increases the system becomes quasibrittle as can be seen from the variations of 
   $\sigma_0(\sigma)$. In Fig. \ref{FIG02}(b) the same variation is plotted for the specific 
   set $s$ = 0.1, $p$ = 0.5 and for four different values of $d$ = 0.1, 0.2, 0.3 and 0.4. Each 
   plot has two regions, one for the first block and the other for the second block. In both the figures 
   $\sigma_0(\sigma)$ varied linearly with $\sigma$ between the two blocks. Here also, as 
   the value of $d$ is increased the system can be seen to change from a brittle phase 
   to a quasibrittle phase. The critical load per fiber of the system is always the maximum value 
   of $\sigma_0(\sigma)$ for all cases. 

      We first study the critical load per fiber $\sigma_c$ for the global failure of the fiber 
   bundle for different values of the parameters $s,d$ and $p$. In particular, we consider 
   the case where the value of the parameter $s$ is fixed at 0.1 and using different values of 
   $p$, we vary the value of the block width $d$ \citep{Smith1982,McCartney1983}. For the 
   estimation of $\sigma_c$ numerically we follow the method of \citep{Smith1981}. We first 
   arrange the breaking thresholds of a particular bundle $\alpha$ in an increasing order 
   such that $b_{(1)}^{\alpha} < b_{(2)}^{\alpha} < b_{(3)}^{\alpha} < ... < b_{(N)}^{\alpha}$. 
   Then, the critical load per fiber $\sigma_c^{\alpha}(N)$ for a particular bundle $N$ can 
   be calculated as  
\begin{equation}
\sigma_c^{\alpha}(N) = \text{max} \Big \{ b_{(1)}^{\alpha}, \frac{N-1}{N}  b_{(2)}^{\alpha}, 
\frac{N-2}{N} b_{(3)}^{\alpha}, ... ,\frac{1}{N} b_{(N)}^{\alpha} \Big\}.
\end{equation}   

      This critical load is then averaged over a considerably large number of configurations to get 
   $<\sigma_c^{\alpha}(N)> = \sigma_c(N)$. We assume that it converges to $\sigma_c(\infty) \equiv \sigma_c$ 
   in the asymptotic limit according to $\sigma_c(N)= \sigma_c + A N^{-1/\nu}$ where $A$ is a 
   constant and $1/\nu$ is a finite size correction exponent. For a sufficiently large $N$ the 
   correction term becomes negligible. In Fig. \ref{FIG03} we have plotted eleven different sets 
   of data for different values of first block probability $p$ tuned between 0 and 1 at equal intervals 
   of 0.1 for $N=2^{18}$. Since this is a considerably large number we expect that this behavior would 
   hold for asymptotically large bundle sizes as well. For small values of $p$ the $\sigma_c$ remains 
   constant in the entire range of variation of $d$. For example, $p=0$ implies all fibers have 
   breaking thresholds larger than $1/2+s$ and the weakest fiber will always have the value $1/2+s$. 
   Since for all values of $d$ the system is always brittle, when the external load per fiber is raised 
   to 0.6, the weakest fiber fails and this leads to a cascade of fiber failures resulting in the break 
   down of the entire fiber bundle. For this reason, the fiber bundle is said to be brittle for this 
   set of parameter values, independent of the block width $d$. As $p$ is increased the $\sigma_c$ 
   decreases and for $p \geq 0.3$ the values are not constant any more. This is because as the number 
   of fibers in the first block increases it lowers the critical load of the system. The same process 
   has also been carried out for $s$ = 0, 0.2, 0.3 and 0.4. The dashed line has the equation 
   $\sigma_c(N,s,p) = 1/2 - s - d$ which means that the set of $(s,p,d)$ for which a $\sigma_c(N,s,p)$ 
   falls on that line is a brittle system.

\section{III. Brittle to quasibrittle transition}

      To describe the brittle to quasibrittle transition we analyze the following three quantities,
   namely: (i) the fraction $f(d,N)$ of fibers broken before the last avalanche, (ii) the average 
   number $\langle M(d,N) \rangle$ of avalanches required for the complete failure of the bundle, 
   and (iii) the average size $\langle \Delta (d,N) \rangle$ of the avalanches \citep{Roy3}. Variations
   of these quantities have been studied against the width parameter $d$ over its entire range. In
   particular, we have estimated the critical value $d_c$ of the width that demarcates the brittle 
   phase of the bundle from its quasibrittle phase.

\subsection{A. Case $p = 0$}  

      First we consider the case when only the right block exists, i.e., $p = 0$ \cite {Pradhan3}. This 
   implies that the breaking thresholds of all the fibers in the bundle are selected in the second block. In 
   this case, the cumulative probability $P_c(b)$ reduces to 
\begin{equation}
P_c(b)=(b-1/2-s)/d 
\end{equation}
   and Eqn. (\ref{EQN04}) becomes
\begin{equation}
\sigma_0(\sigma) = \sigma (1-(\sigma -1/2 -s)/d).
\label{EQN07} 
\end{equation}

\begin{figure}[t]
\includegraphics[width=6.0cm]{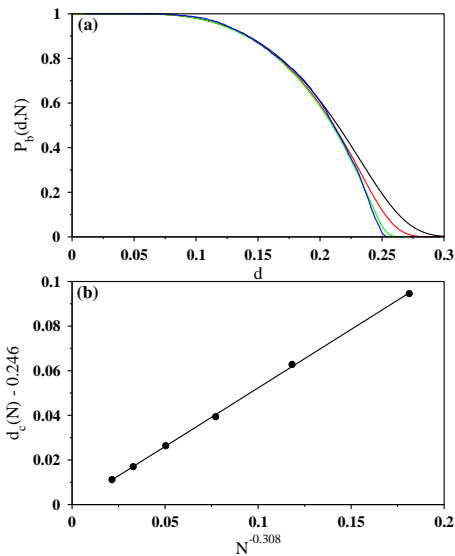}
\caption{
(a) The probability $P_b(d,N)$ that a randomly selected fiber bundle with $s = 0$ and $p = 1$ 
    is brittle, has been plotted against the block width $d$. The bundle sizes used are: 
    $N=2^{8}$(black), $2^{10}$(red), $2^{14}$(green) and $2^{18}$(blue) with $N$ increasing 
    from right to left.
(b) As the bundle size $N$ increases, the critical block width $d_c(N)$ approaches the value
    0.246 as $N^{-0.308}$.
}
\label{FIG05}
\end{figure}
    At the breaking point of the bundle, Eqn. (\ref{EQN07}) has a maximum at $\sigma = \sigma_{0c}(d,s)$ 
    and it is calculated to be 
\begin{equation}
\sigma_{0c}(d,s) = d/2 + (1/2+s)/2.
\end{equation}
   In this situation the value of $\sigma_{0c}(d,s)$ is equal to the minimum value of the breaking 
   thresholds of the bundle, i.e., $1/2+s$. Thus,
\begin{equation}
d_c/2 + (1/2+s)/2 = 1/2+s
\end{equation} 
   which gives $d_c = 1/2+s$, the critical point. This result implies that for a system with all 
   the fibers in the second block, the bundle will always be brittle and no transition can be observed 
   from a brittle to a quasibrittle phase. When $s=0$, the critical width $d_c = 1/2$. This result is
   consistent with the observations of \cite {Pradhan3,Raischel}. In general, for one block uniform
   distribution within the limits of $(b_{min},b_{max})$, the condition for brittleness is: 
   $b_{min} > b_{max}/2$ where, $b_{min}$ and $b_{max}$ are the lower and upper bounds of the 
   breaking threshold distribution.

      Numerically, we study the fraction $f(d,N)$ of fibers broken before the last avalanche against 
   the block width $d$ for four different sizes of the fiber bundle \cite {Roy2}. By definition, $f(d,N)$ is 
   identically zero when the bundle is completely brittle and non-zero when it is quasibrittle. Fig. 
   \ref{FIG04}(a) exhibits the variation of $f(d,N)$ against $d$. As the bundle size $N$ increases, 
   the larger portion of the curve coincides with the $d$ axis. Therefore, the minimal value $d_c(N)$
   of $d$ where $f(d,N)$ is non-zero increases and approaches the value of 1/2. This implies that over
   the entire range of the width parameter $d$, the bundle is in the brittle phase.

\begin{figure}[t]
\includegraphics[width=6.0cm]{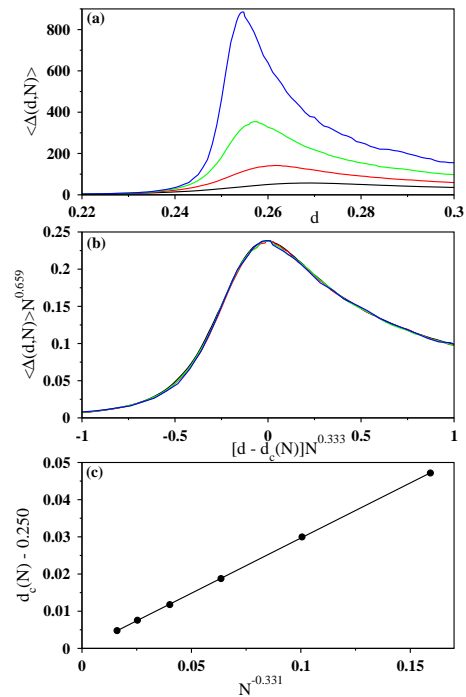}
\caption{
   (a) Plot of the average avalanche size $\langle \Delta (d,N) \rangle$ against the block width $d$.
   (b) The finite size scaling plot of the data in (a). An excellent collapse of the data has been observed
   when $\langle \Delta (d,N) \rangle N^{-0.659}$ has been plotted against $[d-d_c(N)]N^{0.333}$.
   The bundle sizes used in both (a) and (b) are: $N=2^{12}$(black), $2^{14}$(red), $2^{16}$(green) and $2^{18}$(blue)
   with $N$ increasing from bottom to top in panel (a).
   (c) The critical width $d_c(N)$ of the first block obtained from the widths corresponding to the maximum
   values of the average avalanche size $\langle \Delta (d,N) \rangle$ shown in (a). The difference $d_c(N) - 0.250$
   vanishes when plotted against $N^{-0.331}$. Therefore, $d_c(\infty) = 0.250$ is very much consistent
   with the analytical value of $d_c = 1/4$ as given in Eqn. (\ref{EQN12}).}
\label{FIG06}
\end{figure}

      The variation of the scaled average number $\langle M(d,N) \rangle / N $ of avalanches required for the 
   complete failure of the bundle has been plotted against the block width $d$ and is shown in Fig. 
   \ref{FIG04}(b). This quantity is also seen to be increasingly smaller with increasing value of the
   bundle size $N$ indicating the absence of any transition in the system. 

      The size of an avalanche $\Delta(d,N)$ is measured by the number of fibers failed during the 
   avalanche. Following the method of Kun et. al. \cite {Kun} we define the average size of the
   avalanches, excluding the last avalanche. The average avalanche size $\langle \Delta (d,N) \rangle$
   is defined as the ratio of second moment to the first moment of the avalanche sizes, as
\begin{equation}
\langle \Delta (d,N) \rangle = \Sigma_j \Delta^2_j(d,N) / \Sigma_k \Delta_k(d,N)
\label{EQN10}
\end{equation}
   where both the summation indices $j$ and $k$ run over all avalanches except the last avalanche.
   This quantity $\langle \Delta (d,N) \rangle$ has been plotted in Fig. \ref{FIG04}(c) that
   has no maximum for any value of $d$ which proves that for this particular case there is no transition. 
   This result is expected because the case $s = 0$ and $p = 0$ implies that all the fibers are in second block 
   where the bundle always remains in a brittle phase.       

\begin{figure}[t]
\includegraphics[width=6.0cm]{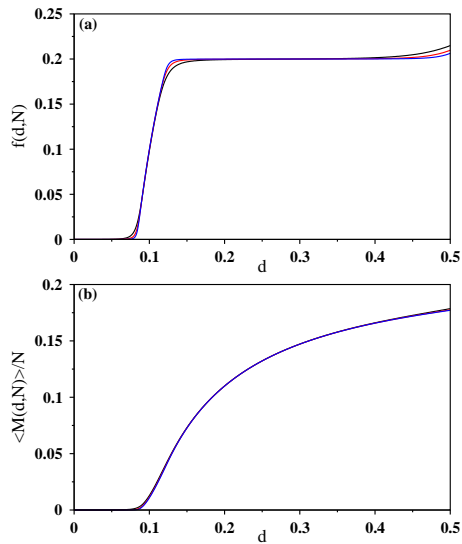}
\caption{Variation of the same quantities plotted in Fig. \ref{FIG04}(a) and Fig. \ref{FIG04}(b), 
i.e., $f(d,N)$ and $\langle M(d,N) \rangle / N$ against the block width $d$ have been displayed 
for $s=0$ and $p = 0.2$. The bundle sizes are $N=2^{12}$ (black), $2^{14}$ (red) and $2^{16}$ (blue). 
These plots are characteristically different from those in Fig. \ref{FIG04}. 
Each indicates the existence of a transition from the brittle phase to the quasibrittle phase.}
\label{FIG07}
\end{figure}

\subsection{B. Case $p = 1$}

The case with $p = 1$ implies that all the fibers in the bundle have their breaking 
thresholds drawn from the first block. For this case 
\begin{equation}
\sigma_{0c}(d,s) = d[1+ (1/2 -s -d)/d]/2.
\end{equation}
On equating $\sigma_{0c}(d,s)$ to the value of the lowest breaking threshold for 
this case $1/2-s-d$ we get the transition point as
\begin{equation}
d_c = (1/2-s)/2.
\label{EQN12}
\end{equation}
Thus, when all the fibers in the bundle are in the first block then the value of 
$d_c$ decreases linearly with increase in $s$. 
\begin{figure}[t]
\includegraphics[width=6.0cm]{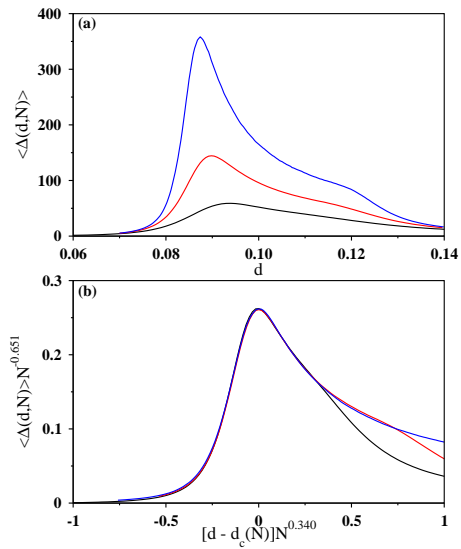}
\caption{
   (a) Plot of the variation of the quantity plotted in Fig. \ref{FIG04}(c), i.e., $\langle \Delta(d,N) \rangle / N$ 
   against the block width $d$ have been displayed for $s=0$ and $p = 0.2$ for $N=2^{12}$(black), $2^{14}$(red), 
   and $2^{16}$(blue) with $N$ increasing from bottom to top. The quantity shows a maximum at a certain value of 
   $d_c$ which indicates a transition from a brittle to a quasibrittle state.
   (b) Finite size scaling of the average avalanche size $\langle \Delta(d,N) \rangle N^{-0.651}$ against 
   $[d-d_c(N)]N^{0.340}$ using the data in (a) exhibits an excellent data collapse.}
\label{FIG08}
\end{figure}

      In a specific case, the result of $d_c = 1/4$ from Eqn. (\ref{EQN12}) for $s=0$ and $p = 1$ has been 
   verified numerically. The probability $P_b(d,N)$ that a randomly selected sample of 
   the fiber bundle is brittle has been plotted against $d$ in Fig. \ref{FIG05}(a) for four
   different bundle sizes. The critical width $d_c(N)$ for a specific bundle size $N$
   is defined as the minimum value of $d$ for which $P_b(d,N)$ vanishes. We assume the transition 
   to be continuous and that it follows the usual finite size scaling relations. The estimated 
   values of $d_c(N)$ are assumed to converge to their asymptotic value $d_c$ as:
\begin{equation}
d_c(N) - d_c = B N^{-1/\nu}. 
\end{equation}
   To estimate the asymptotic value $d_c$ and the exponent $\nu$ we have plotted the $d_c(N)$ 
   against $N^{-1/\nu}$ in Fig. \ref{FIG05}(b). The precise value of the exponent is tuned so 
   that we get the best straight line with minimal fitting error. Our best estimate from this 
   plot are $d_c = 0.246$ and $1/\nu = 0.308$.
   
      The critical width $d_c(N)$ has also been estimated from the statistics of avalanche sizes.
   The average size of the avalanches $\langle \Delta (d,N) \rangle$ given by Eqn. (\ref{EQN10}) 
   has been studied and plotted against $d$ in Fig. \ref{FIG06}(a) for four different values of 
   the bundle sizes. It is seen that for every bundle size the curve has a maximum at a certain value
   of $d$ which we assume as the second definition of $d_c(N)$. A finite size scaling of the 
   data turned out to be very nice when we plotted $\langle \Delta (d,N) \rangle N^{0.659}$ against
   $[d - d_c(N)]N^{0.333}$. We use the $d_c(N)$ values estimated from the peak positions and 
   in Fig. \ref{FIG06}(b) all four curves fall very closely on one another. In Fig. \ref{FIG06}(c)
   we again plot $d_c(N) - d_c$ against $N^{-1/\nu}$ and tune $\nu$ to get the best fitted straight
   line. Our results are $d_c = 0.250$ and $1/\nu = 0.331$. 

\begin{figure}[t]
\includegraphics[width=6.0cm]{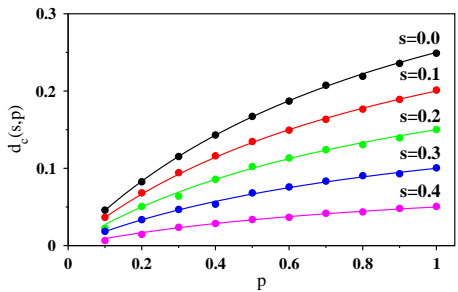}
\caption{
      The extrapolated values of the critical block width $d_c(s,p)$ in the asymptotic limit of large 
   bundle sizes have been plotted using filled circles against the first block probability $p$ for 
   different values of the separation parameter $s = 0$ (black), 0.1 (red), 0.2 (green), 0.3 (blue) 
   and 0.4 (magenta) with $s$ increasing from top to bottom. The continuous curves are the plots of 
   the Eqn. (\ref{EQN15}) which match very well with the numerical data.}
\label{FIG09}
\end{figure}

\subsection{C. Case $0 < p < 1$}

       We have further observed that for other intermediate values of the first block probability 
   parameter $p$, again with the separation parameter $s = 0$, there exists non-trivial phase 
   transitions from the brittle to the quasibrittle phases. For example, in a particular case of 
   $p = 0.2$, we have again studied the same quantities, namely the fraction $f(d,N)$ of fibers 
   broken before the last avalanche, the average number $\langle M(d,N) \rangle$ of avalanches 
   required for the complete failure of the bundle and the average size $\langle \Delta(d,N) \rangle$ 
   of the avalanches \cite {Danku}. These quantities have been plotted against the block width $d$ in Figs. \ref{FIG07}(a), 
   \ref{FIG07}(b) and \ref{FIG08}(a) where $d$ has been tuned from 0 to $1/2$ at the interval of 
   0.0001. It is again observed that $\langle \Delta(d,N) \rangle$ exhibits a maximum at $d_c(N)$ 
   and grows with the bundle size indicating a phase transition. Finally, in Fig. \ref{FIG08}(b) 
   we have plotted the scaled variable $\langle \Delta(d,N) \rangle N^{0.651}$ against 
   $[d - d_c(N)]N^{0.340}$ which exhibits a good collapse of data. The value of $d_c$ for large 
   $N$ have been estimated by extrapolating the bundle sizes against $N^{-0.325}$ over 
   $N=2^{14},2^{16}$ and $2^{18}$. 

      Similarly, the calculation for $d_c(s,p)$ for $s = 0$ has been repeated for the other values 
   of $p$ in the range $0 < p \le 1$ at the interval of 0.1 and plotted in Fig. \ref{FIG09}. 
   Moreover, four other sets of data of $d_c(s, p)$ against $p$ for $s =$ 0.1, 0.2, 0.3
   and 0.4 have been plotted in the same Fig. \ref{FIG09}. For a particular value 
   of $s$, $d_c$ is observed to increase with increase in $p$. 
   This is because as the number of fibers in the first block increases, the load accumulated
   during the breaking process increases leading to the increase in the
   possibility of breaking the weakest fiber from the second block.
  
      From Eqn. (\ref{EQN04}), one obtains the maximum value of $\sigma_{0c}(d,s,p)$ for a 
   quasibrittle state for the first block as
\begin{equation}
   \sigma_{0c}(d,s,p) = d [1+p(1/2 -s-d)/d ]/(2p).
\end{equation}

      When a bundle is in a brittle state, the critical load per fiber $\sigma_{0c}(d,s,p)$ at the 
   breaking point of the bundle should be equal to the value of the weakest fiber in it i.e. 
   ($1/2-s-d$). Therefore, on solving for $d$ we get
\begin{equation}
d_c(s,p) = p(1/2-s)/(1+p).
\label{EQN15} 
\end{equation}
   Substituting the values of $s=0$ and $p=1/2$ one gets back the well established result of the 
   critical width $d_c=1/6$ \cite{Subhadeep}. The cases $p=0$ and $p=1$ gives back the 
   results of the limiting cases discussed in previous sections.
   The Eqn. (\ref{EQN15}) has been plotted in Fig. \ref{FIG09} for $s = 0, 0.1,0.2,0.3$ and $0.4$ along 
   with the numerical results.

\begin{figure}[t]
\includegraphics[width=6.0cm]{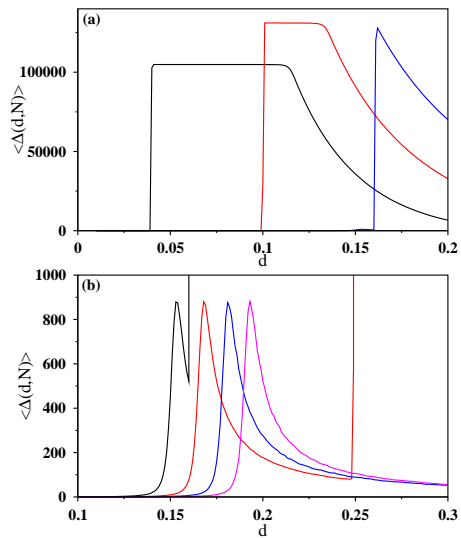}
\caption{
   Plot of the average avalanche size $\langle \Delta(d,N) \rangle$ against the block 
   width $d$ for different values of (a) $p$ = 0.4, 0.5, 0.6 and (b) $p$ = 0.6, 0.7, 0.8, 
   0.9 for a bundle of size $N=2^{18}$ with $s = 0.1$. Here, $p$ increases from left to right in 
   both the figures.}
\label{FIG10}
\end{figure}

      Next, we studied the average avalanche size $\langle \Delta(d,N) \rangle$ against $d$. 
   The average avalanche size for cases when $s \neq 0.0$ behaves differently depending on the 
   fraction of fibers $p$ in the first block. This quantity is plotted in Fig. \ref{FIG10}(a) 
   for $s$ = 0.1 and for different values of $p$ = 0.4,0.5 and 0.6. For small values of $d$
   for $s = 0.1$ and $p = 0.4$ the value of $\langle \Delta(d,N) \rangle$ is vanishingly small
   followed by a discontinuous jump leading to a plateau. Then it's value sharply decreases as 
   $d$ is increased further. Similar curves are observed for $p$ = 0.5 as well. However, no
   plateau is observed for $p$ = 0.6. On the other hand for a much smaller value of 
   $\langle \Delta(d,N) \rangle$, a small peak is observed for the same plot as shown in Fig. 
   \ref{FIG10}(b). Such behavior is observed till $p$ = 0.7 after which only the small peaks 
   remain and the large peaks vanish.

      The formation of the plateau region occurs because when $d$ is small, the bundle is
   in the brittle regime and all the fibers from the first block break in either one or only
   a few avalanches.
   Thus the average avalanche size $\langle \Delta(d,N) \rangle$ excluding the last avalanche remains constant 
   as the total load released by these broken fibers is not enough to break even the weakest fiber 
   in the second block. At the edge of the plateau the number of avalanches increase significantly and 
   thus the value of $\langle \Delta(d,N) \rangle$ is seen to fall sharply. In this case, we define 
   the critical width $d_c$ to be located at the end of the plateau instead of the beginning. This 
   is because even though more than one avalanche is required to break all the fibers in the first block, 
   the breakdown is rapid and the number of such avalanches is very small.

      The value of the width $d$ at the right edge of the plateau where $\langle \Delta(d,N) \rangle$
   sharply decreases is considered to be the critical width $d_c(N)$ for the system size $N$. The 
   numerical values obtained have been plotted in Fig. \ref{FIG09}. For $s$ = 0.1 and $p$ = 0.6 and 0.7, two significant
   peaks have been observed. The value of $d$ at which the smaller peak occurs as shown in 
   Fig. \ref{FIG10}(b) has been defined as the $d_c(N)$ for these cases. This is because 
   the small peak indicates that a considerable number of avalanches occur of small sizes as $p$ is 
   large enough and $d$ is not too small. All the values of $d_c(N)$ obtained through the above 
   mentioned method have been observed to match very closely with the analytical result obtained in 
   Eqn. (\ref{EQN15}).  
  
\section{IV. Summary}

      To summarize, we have studied the brittle to quasibrittle transition in a compound fiber bundle 
   model characterized by bimodal distribution of fiber breaking thresholds. We have observed that the critical load
   per fiber for the failure of the bundle strongly depends on all the three parameters, namely, the width $d$ of the 
   blocks, the separation $s$ between the blocks and the probability $p$ of the first block. We have 
   parameterized such a transition using three different quantities, namely: 
   (i)   the average fraction of fibers broken before the last avalanche, 
   (ii)  the average number of avalanches required for the complete breakdown of a fiber bundle and 
   (iii) the average avalanche size excluding the last avalanche. 
   In addition, we could formulate a general expression for the critical width $d_c(s,p)$ of the 
   phase transition analytically and have verified it by the numerical analysis. 

      We thank Mr. Sumanta Kundu very much for many helpful discussions.

\end {document}